\documentstyle[12pt]{article}

\def\<{\langle} \def\>{\rangle}  
  \def\Tr{{\rm Tr}}
\def\half{{\scriptstyle{ 1\over 2}}}  
     
  \def\dag{^\dagger} 
 \def\hs1{\hskip1mm} \def\h10{\hskip10mm}
\def\ket#1{|#1\rangle} \def\bra#1{\langle#1|}
\def\expect#1{\langle#1\rangle} \def\Qhat{{\hat Q}} \def\Phat{{\hat P}}
\def\Hhat{{\hat H}} \def\Ohat{{\hat O}} \def\Lhat{{\hat L}}
\def\Ldag{{\hat L}^\dagger} \def\deltaQT{{\Delta\Qhat}^2}
\def\deltaPT{{\Delta\Phat}^2} \def\deltaQP{\Delta\Qhat\Delta\Phat}
\def\deltaPQ{\Delta\Phat\Delta\Qhat} 
 \def\Proj{{\hat {\cal P}}}
\def\ahat{{\hat a}} 

\begin{document}

\parskip 4mm plus 1mm \parindent=0pt

\title{Quantum chaos in open systems: \\
a quantum state diffusion analysis}

\author{Todd A. Brun, Ian C. Percival, and R\"udiger Schack \\
Department of Physics \\
Queen Mary and Westfield College, University of London \\
Mile End Road, London E1 4NS, England}

\date{\today}

\maketitle

 \hfil

\begin{abstract}
Except for the universe, all quantum systems are open, and according
to quantum state diffusion theory, many systems
localize to wave packets in the neighborhood of phase space points.
This is due to decoherence from the
interaction with the environment, and
makes the quasiclassical limit of such systems both more realistic
and simpler in many respects than the more familiar quasiclassical limit
for closed systems.  A linearized version of this theory leads to
the correct classical dynamics in the macroscopic limit,
even for nonlinear and chaotic systems.  We apply the theory to the
forced, damped Duffing oscillator, comparing the
numerical results of the full and linearized equations, and argue that
this can be used to make explicit calculations in the decoherent histories
formalism of quantum mechanics.
\end{abstract}

\vfil

QMW Theory Preprint QMW-PH-95-33

\eject

\section{Introduction}

One of the chief questions in quantum theory is how classical behavior
arises in quantum mechanical systems.  A na\"\i ve description
of a closed system, using the Schr\"odinger equation,
very quickly loses all resemblance to classical mechanics.  Even a highly
localized wave packet can, given sufficient time, spread until it
occupies all accessible regions of phase space.  Such delocalized
states are highly non-classical, and quite unlike what we
normally observe.

However, if there is any significant interaction with the environment,
such as a measurement, the system becomes open, and the Schr\"odinger
equation is no longer valid.  Macroscopic systems, with the exception of
the entire universe, are never completely isolated.
Even in interstellar space, interactions with thermal
background radiation and encounters with stray atoms
prevent macroscopic superpositions from arising \cite{JoosZeh}.

The master equation approach has long been used to treat
open quantum systems.  In it, one averages
over the unknown effects of interaction with the environment to
compute the probabilities of possible output states.  This has been used
in an enormous range of experimental set-ups.

Modern experiments, however, often enable one to follow the behavior of a
single quantum system as it evolves, influenced (but not destroyed)
by the environment, including the measurement apparatus itself.
Quantum state diffusion (QSD) provides a formalism which can be used to
describe such situations \cite{GisPer}.
In it, a system is always in a pure state,
evolving according to a stochastic nonlinear Langevin-It\^o equation.
The equation includes both collective effects of the environment
(such as dissipation and
localization) and random fluctuations.  For most open systems, the classical
limit is qualitatively very different than for closed systems.  QSD shows this
particularly clearly, with its dynamical localization producing wave
packets which follow approximately classical trajectories.  In this sense,
the classical limit of open systems is far simpler than that for closed
systems.

In section 2 we describe the
QSD formalism, and compare it with the decoherent histories approach
to quantum mechanics \cite{Zurek,Griffiths,Omnes,GMHart1,GMHart2,GMHart3}.
It has recently been shown that these two approaches
are related, and in fact provide complementary views of systems interacting
with environments \cite{DGHP}.

In section 3 we examine the quantum dynamics of linear systems and
their classical limit, and how nonlinear systems can
also be treated in this limit.  We show that this is
very different for open systems with localization than for
closed systems.  We briefly outline the importance
of localization in efficient numerical simulation of QSD, and
how it gives QSD remarkable advantages over other simulation methods
for open systems.  This builds on the results
in \cite{Schack1}, for a particular example.

In section 4 we
examine the problem of quantum dissipative chaos.  While
considerable work has been done on quantum equivalents to classical
Hamiltonian chaotic systems, very little has been done for dissipative
chaos.  In part, this is due to the challenging problems involved in
studying dissipation in quantum systems.
Spiller and Ralph \cite{SpillRalph} have shown how quantum
state diffusion can be used to model dissipative chaotic systems.
We examine a particular model,
the forced, damped Duffing oscillator, to illustrate both the classical
and quantum characteristics of dissipative chaos, and argue the
relevance of these results to treatments using decoherent histories.
The theoretical arguments are backed up by numerical results.

Finally, in section 5 we re-examine our results, and draw conclusions
on the usefulness of QSD both as a theoretical and a practical tool in
studying the quasiclassical limit of open quantum systems.

\section{Quantum dynamics}

\subsection{Quantum state diffusion (QSD)}

The master equation is used in the Markov approximation.
It can then be expressed in {\it Lindblad form} \cite{Lindblad}:
\begin{equation}
\dot\rho = - {i\over\hbar}[\Hhat,\rho]
  + \sum_{j=1}^m \biggl( \Lhat_j \rho \Lhat_j\dag
  - {1\over2} \Lhat_j\dag \Lhat_j \rho 
  - {1\over2} \rho \Lhat_j\dag \Lhat_j \biggr),
\label{master_equation}
\end{equation}
where $\rho$ is the density operator,
$\Hhat$ is the Hamiltonian and the $m$
operators $\Lhat_j$ are the {\it Lindblad operators},
which model the effects of
the environment.  Different choices of these $\Lhat_j$ have been used to
simulate a wide range of physical situations, including measurements.

The evolution of a single open system is unpredictable.
While an {\it ensemble} of systems may evolve in
a perfectly deterministic manner, any {\it single} experimental run must have a
strong stochastic component.  There are many ways of ``unraveling'' the
master equation into trajectories; the
quantum state diffusion (QSD) formalism is one of these, with unique symmetry
and localization properties \cite{GisPer,Percival1}.

In QSD, a state evolves according to the It\^o equation
\begin{eqnarray}
\ket{d\psi} = && - {i\over\hbar} \Hhat \ket\psi dt
  + \sum_{j=1}^m \biggl( \<\Lhat_j\dag\> \Lhat_j
  - {1\over2} \<\Lhat_j\dag\>\<\Lhat_j\>
  - {1\over2} \Lhat_j\dag \Lhat_j \biggr) \ket\psi dt \nonumber\\
&& + \sum_{j=1}^m ( \Lhat_j - \< \Lhat_j \> ) \ket\psi d\xi_j\;,
\label{qsd_equation}
\end{eqnarray}
where the $d\xi_j$ are $m$ independent
complex random differential variables
representing a complex Wiener process.  They obey relationships
\begin{equation}
M(d\xi_i) = M(d\xi_i d\xi_j) = 0,\ \ M(d\xi_i d\xi_j^*) = dt\delta_{ij},
\end{equation}
where $M$ represents the ensemble average.

This unraveling of the master equation has many advantages.  It is invariant
under unitary transformations among the Lindblad operators, an invariance shared
with the master equation, and which is unique to QSD.
By taking the ensemble mean of the projector
$\ket\psi \bra\psi$, one reproduces the master equation above:
$M(\ket\psi \bra\psi) = \rho$.  This gives a very simple interpretation for
the relationship between the trajectories $\ket{\psi(t)}$
and the density operator
$\rho$:  $\rho(t)$ is the ensemble of all possible states
$\ket{\psi(t)}$; over the
course of many runs of an experiment, the $\ket\psi$'s will tend to reproduce
the density operator $\rho$, while in a single run, the system is in a
state $\ket\psi$ (which depends on the choice of unraveling).
Thus, expectation values
$\expect{\hat O}_\rho$ calculated with a
density operator $\rho$ would include both the quantum expectation value
$\expect{\hat O}_\psi$ and the ensemble mean over all the $\ket\psi$s.

Numerically there is often a large advantage in representing a system
by a state rather than a density operator.  If it requires $N$ basis states
($2N$ real numbers) to represent the wavefunction,
the density operator will require $N^2$ real numbers.  Thus, even
though many runs may be required to produce an accurate answer, the factor
of order $N$ saved on each individual run can still result in large savings
of computer time and memory.
This is particularly true if one takes advantage of QSD's localization
properties \cite{Schack1}.

\subsection{Decoherent histories}

An extremely promising interpretation of quantum mechanics
is the decoherent (or
consistent) histories formalism, developed by Griffiths, Omn\`es, and
Gell-Mann and Hartle \cite{Griffiths,Omnes,GMHart1,GMHart2,GMHart3}.
In this formalism, quantum systems are described
by sets of possible histories $\{\alpha\}$.
These histories 
ascribe certain definite values, or ranges of values, to selected
variables at successive points in time.  Along with the histories, there
is a {\it decoherence functional} $D[\alpha,\alpha']$ on pairs of histories.
For a set of histories $\{\alpha\}$ to consistently
describe the quantum system, they must satisfy the {\it decoherence
criterion}
\begin{equation}
D[\alpha,\alpha'] = \delta_{\alpha \alpha'} p_\alpha,
\label{decoherence}
\end{equation}
where $p_\alpha$ is the probability for the history $\alpha$ actually to
occur.  If the decoherence criterion is met, then these histories obey
the usual classical probability sum rules; one can
assert that one or another history actually occurred, without the
kind of interference effects seen in the two-slit experiment.
In the case of Schr\"odinger's cat, one history would include a live cat
and another a dead cat, without reference to observers or measurements.

In standard non-relativistic quantum mechanics, the usual choice of
histories consists of an
exhaustive set of orthogonal projection operators
$\{\Proj^i_{\alpha_i}(t_i)\}$ at each time of interest $t_i$.
\begin{equation}
\sum_{\alpha_i} \Proj^i_{\alpha_i}(t_i) = {\hat 1},\ \ 
  \Proj^i_{\alpha_i}(t_i) \Proj^i_{\alpha'_i}(t_i) =
  \delta_{\alpha_i \alpha'_i} \Proj^i_{\alpha_i}(t_i).
\end{equation}
These projections
represent different alternatives at the time $t_i$.  A single history
$\alpha$ represents a choice of one $\Proj^i_{\alpha_i}(t_i)$ at each
time $t_i$,
represented by the string of projection
choices $\alpha_i$.

The decoherence functional can be written
\begin{equation}
D[\alpha,\alpha'] = \Tr \biggr\{ \Proj^n_{\alpha_n}(t_n) \cdots
  \Proj^1_{\alpha_1}(t_1) \rho \Proj^1_{\alpha'_1}(t_1) \cdots
  \Proj^n_{\alpha'_n}(t_n) \biggl\},
\label{decoherence_functional}
\end{equation}
and a set of histories must satisfy (\ref{decoherence}).

One might be tempted to choose as a set of histories the precise values
of some complete set of variables at each point in time.  For example,
for a single particle one might have its histories be its different
possible trajectories $x(t)$.  Histories of this sort almost never
decohere.  In general, a fair degree of {\it coarse-graining}
is required to even approximately satisfy (\ref{decoherence}).
Since finding exactly decoherent sets of histories is
difficult, the usual course is to find a set of histories which
satisfies the decoherence criterion to a finite level of accuracy
\cite{DowkHall}.
It is conjectured that there is an exactly decoherent set of histories in a
neighborhood of every such approximately decoherent set
\cite{DowkKent}.

One simple (and therefore popular) type of coarse-graining is to
separate the system into a distinguished subsystem and an environment
\cite{Zurek,GMHart3,Brun1}.
Tracing over the environment then produces a set of coarse-grained
histories in terms of the ``distinguished variables'' alone.  It has been
shown that under a wide variety of circumstances, histories coarse-grained
in this way approximately decohere, with increasing precision as
the distinguished system approaches the classical limit \cite{Zurek}.

This type of coarse-graining brings us close to the situation
of interest for quantum state diffusion.  It has recently been shown
that for a large class of open quantum systems
there is a close connection between quantum state diffusion
and decoherent histories \cite{DGHP}.
The different possible
QSD trajectories correspond to alternative histories; these
histories approximately decohere, and occur with the
same probabilities as in the decoherence formalism.  In this context
one can think of QSD as a dynamical form of the decoherent
histories formalism, or decoherent histories as a global view of QSD.

\section{Open quantum systems and the quasiclassical limit}

\subsection{Linear quantum systems}

An important class of systems which can be treated with the QSD
formalism are {\it linear systems}, with a Hamiltonian function
quadratic in $\Qhat$ and $\Phat$, and Lindblad operators given by a (not
necessarily Hermitian) combination
$c_q \Qhat + c_p \Phat$.  These systems have
been treated at length by a number of researchers \cite{Diosi1,SalGis}.

Halliwell and Zoupas \cite{HallZoup} have shown that these linear
systems tend to localize with time onto Gaussian wave packets in phase
space.  These wave packets follow paths in phase space which are
close to paths defined by the classical (or quasiclassical) equations
of motion.

From the QSD equation (\ref{qsd_equation})
the It\'o equation for the
expectation value of an observable is
\begin{eqnarray}
& d\expect\Ohat = \bra{d\psi}\Ohat\ket{\psi} +
  \bra{\psi}\Ohat\ket{d\psi} + \bra{d\psi}\Ohat\ket{d\psi} \nonumber\\
& = - i \expect{[\Ohat,\Hhat]} dt + \expect{\Ldag\Ohat\Lhat -
  \half\{\Ldag\Lhat,\Ohat\} } dt +
  ( \expect{\Ohat\Lhat} - \expect\Ohat \expect\Lhat ) d\xi + \nonumber\\
& ( \expect{\Ohat\Ldag} - \expect\Ohat \expect\Ldag ) d\xi^*.
\end{eqnarray}
If the Hamiltonian is of the usual form
\[
\Hhat = \Phat^2/2m + V(\Qhat),
\]
and the Lindblad operator $\Lhat$ is linear
then we can readily evaluate this expression for simple operators $\Ohat$.

Halliwell and Zoupas have shown that for linear systems one can, to a
very good approximation, reduce the QSD equation to a set of coupled
equations for $\expect\Qhat$, $\expect\Phat$, $\expect\deltaQT$,
$\expect\deltaPT$, and $\expect{\deltaQP + \deltaPQ}$.
So long as the wave packet remains
localized, so that the higher moments can be neglected,
the QSD evolution can be reduced to these five
equations.  The equations for $\expect\Qhat$ and $\expect\Phat$
correspond to the equations of motion for a
classical path in phase space, modified by the presence of noise and
small diffusive corrections which depend on the higher moments.

This is similar to the Ehrenfest theorem for the Schr\"odinger
equation, in which it can be shown that wave packets localized in phase
space follow approximately classical paths.  Unlike the Ehrenfest theorem,
however, in QSD this localization usually becomes better and better with
time, rather than the reverse
\cite{Percival1,HallZoup}.

\subsection{Nonlinear quantum systems}

While the equations of Halliwell and Zoupas were derived
for linear systems, they
provide a good approximation for the classical limit even of nonlinear
systems.  In this classical limit, the width of
a localized wave packet is small compared to the spatial variance
of the potential (and its higher derivatives).  Thus, to the wave packet
the potential seems locally quadratic, though slowly time-varying, and we
would expect many of these localization results to apply.

In the classical limit, therefore, one can treat nonlinear problems
by linearizing about the quasiclassical path, and solving for the
moments of the wave packet.  This is quite similar to the classical
technique of linearizing about a known orbit to examine the behavior
of nearby solutions.

Halliwell and Zoupas
\cite{HallZoup} have shown that
solutions tend in the mean towards localized Gaussian
wave packets when the third-derivative and higher terms in the
potential are negligible, i.e., when
\begin{equation}
\expect{V'(\Qhat)} \approx V'(\expect\Qhat),
\end{equation}
\begin{equation}
\expect{\Qhat V'(\Qhat)} - \expect\Qhat \expect{V'(\Qhat)}
 \approx \expect\deltaQT V''(\expect\Qhat),
\end{equation}
and
\begin{equation}
\expect{\Phat V'(\Qhat) + V'(\Qhat) \Phat}
  - 2 \expect\Phat \expect{V'(\Qhat)} \approx
  \expect{\deltaQP+\deltaPQ} V''(\expect\Qhat).
\end{equation}

Numerical investigation \cite{GisPer,Schack1}
as well as theoretical arguments \cite{Percival1}
indicate that localization occurs in almost all systems, whether
linear or not.  If one thinks of QSD in terms of decoherence this
is not surprising.  Many people have shown that decoherence
through interaction with an environment is extremely rapid.  This
corresponds to superpositions quickly reducing to highly localized
states.  In most cases, this localizing effect of the
environment predominates over dispersive effects due to quantum spreading.

\subsection{Localization and the moving basis}

One practical consequence of localization is that it is possible, by
choosing an appropriate basis, to reduce significantly the number of
basis states needed to represent the wave packet \cite{Schack1}.
If a wave packet
is localized about a point $(q,p)$ in phase space, it might require a
great many of the usual number states $\ket{n}$ to represent it.  If
instead we represent the wave packet in terms of {\it excited coherent
states} $\ket{q,p,n} = {\hat D}(q,p)\ket{n}$, relatively few basis states
are required, with obvious savings in terms of computer storage and
computation time.  Here, ${\hat D}(q,p)$ is the coherent state
displacement operator,
\begin{equation}
{\hat D}(q,p) = \exp {i\over\hbar} \biggl( p{\hat Q} - q{\hat P} \biggr),
\end{equation}
which displaces the excited state $\ket{n}$ to the excited coherent state
$\ket{q,p,n}$ centered on $(q,p)$.
This representation of the state by
classical $(q,p)$ and quantum $\ket{q,p,n}$ is called the
{\it moving basis} or {\it mixed} representation,
or MQSD, and has been discussed extensively elsewhere
\cite{Percival2,Schack1}.

\section{The Duffing oscillator}

\subsection{Quantizing the forced damped Duffing oscillator}

A good example of a nonlinear system is the forced, damped
Duffing oscillator.
This has a classical equation of motion
\begin{equation}
{d^2x\over dt^2} + 2\Gamma{dx\over dt} + x^3 - x = g\cos(t),
\label{duffing_eom}
\end{equation}
and for some choices of $\Gamma$ and $g$ is chaotic
\cite{Gutzwilleretal}.

Because the equation of motion includes explicit time-dependence, the
solutions lie in a three-dimensional phase space ${x,p,t}$.  It is
helpful to consider a discrete surface of section of this system.
Let $(x_0, p_0)$ be the initial point of the forced, damped Duffing
oscillator at time $t_0=0$.  Then we can define a {\it constant phase map}
in the $x$-$p$ plane by the sequence of points $(x_n, p_n) = (x(t_n), p(t_n))$
at times $t_n = 2\pi n$.  Figure 1 illustrates this in the
chaotic regime, where we can clearly see from the surface of section the
fractal structure of the strange attractor.

Quantizing the Duffing oscillator is straightforward using the QSD formalism.
The Hamiltonian operator is
\begin{equation}
\Hhat(\Qhat,\Phat,t) =
  \Phat^2/2m + \Qhat^4/4 - \Qhat^2/2 + g\cos(t) \Qhat
  + \sqrt\Gamma (\Qhat\Phat + \Phat\Qhat),
\end{equation}
and the damping is represented by a Lindblad operator
\begin{equation}
\Lhat = 2\sqrt\Gamma \ahat = \sqrt{2\Gamma}( \Qhat + i\Phat),
\end{equation}
where we have assumed $\hbar = 1$.
The last term in the Hamiltonian is an ansatz, added to give the correct
equations of motion in the classical limit; it is necessary due to the
simplistic model of the dissipative environment.

This system is far from classical.  To
go to the classical limit, we introduce a scaling factor $\beta$,
\begin{equation}
\Hhat_\beta(\Qhat,\Phat,t) = \Phat^2/2m + \beta^2 \Qhat^4/4 - \Qhat^2/2
  + (g/\beta)\cos(t) \Qhat + \sqrt\Gamma (\Qhat\Phat + \Phat\Qhat).
\label{hamiltonian}
\end{equation}
As we reduce $\beta$, the scale
of the problem (compared to $\hbar$) increases by $1/\beta$ in $x$ and $p$,
without altering the classical dynamics.  Thus,
$\beta \rightarrow 0$ is the classical limit of this system.  Classical behavior
should emerge from the system in this limit.  This is
supported by the numerical calculations. (See section 4.3.)

\subsection{Application to decoherent histories}

The forced, damped Duffing oscillator has been treated elsewhere using
the decoherent histories formalism \cite{Brun2,Brun3,Brun4}.
While this approach has great
theoretical power, it does not lend itself readily to numerical solution,
which is generally necessary for this kind of nonlinear problem.
Fortunately, as it has been shown that QSD is very closely related to
a certain kind of coarse-grained decoherent history \cite{DGHP},
one can use the
numerically efficient QSD method to perform a quantum Monte Carlo
calculation for open systems of this sort.

It has been argued \cite{DowkHall,GMHart3,Brun1,Brun2} that
solutions of these open systems should peak about the quasiclassical
equations of motion, i.e., they should behave in the classical limit
like a classical problem with noise.  This is amply demonstrated by
the numerical results in section 4.3.

By choosing an appropriate set of decoherent histories one can find a
quantum equivalent to the Poincar\'e surface of section defined by
(\ref{duffing_eom}) above.  Consider the times $t_i = 2\pi i$, with a
set of projection operators $\Proj_{q_i,p_i}(t_i)$ onto local cells of
phase space defined at each time, these cells centered at $(q_i,p_i)$.
Although there are no true
orthogonal projections onto cells of phase space, there are a number
of approximate projections; the simplest are the coherent state
projectors $\Proj_{qp} = \ket{q,p}\bra{q,p}$,
and there are many others \cite{Omnes,Halliwell1,Twamley,Brun2,Anasto}.
We can write down
the decoherence functional
\begin{equation}
D[\{q_i,p_i\},\{q_i',p_i'\}] =
  \Tr\biggl\{\Proj_{q_n,p_n}(t_n) \cdots \Proj_{q_1,p_1}(t_1) \rho
  \Proj_{q_1',p_1'}(t_1) \cdots \Proj_{q_n',p_n'}(t_n) \biggr\},
\label{D_phase}
\end{equation}
where $\rho = \ket\psi\bra\psi$ is the initial density operator,
for simplicity taken to be pure.

It has been shown that in systems which interact strongly with an
environment, the off-diagonal terms of the decoherence functional
decay very rapidly.  So long as the time between projections
$t_{i+1} - t_i$ is large compared to the decoherence time $t_D$,
and the phase space cells are large compared to $\hbar$,
the histories described by (\ref{D_phase}) will decohere to a good
level of precision.
In the quasiclassical
limit, the probabilities $D[\{q_i,p_i\},\{q_i,p_i\}]$ should
be sharply peaked about the classical solution.

As shown by Di\'osi,
Gisin, Halliwell and Percival \cite{DGHP},
quantum state diffusion trajectories
will give the same probabilities as the decoherence functional for
histories which project onto the same quantities (in this case coherent
states in phase space) and are at least as coarse-grained as the level of
QSD localization.  By following a QSD trajectory, therefore, and looking
at the expectation values of $\expect\Qhat$ and $\expect\Phat$ at
the times $t_i$, one has picked out
a single decoherent history with the correct probability.

\subsection{Numerical example}

We solved the forced, damped Duffing oscillator
numerically by integrating the QSD equation,
both in full (using MQSD) and in the linearized approximation
of Halliwell and Zoupas,
using a C++ quantum simulation package we have developed
\cite{Schack1,Schack2}.
All of these results were generated on standard
486 PC's running LINUX.

The classical unscaled problem is bounded within a small
region of phase space.  (See figure 1.)
Clearly, the quantized version
of this problem (with $\hbar=1$) should be far from the classical limit.
One would expect to observe little trace of the classical fractal
structure.  The expectation values of $\Qhat$ and $\Phat$ should be
dominated by noise.

Examining figure 2a, we see that the results of the
numerical calculation match our expectations very well.  The expectation
values appear randomly distributed; they are dominated almost
completely by the stochastic terms of the equation.

Approaching the classical limit $\beta\rightarrow0$,
more and more of the classical structure of the attractor appears.
At first the broad outlines of the attractor are formed, then increasing
levels of substructure.  (Figures 2b--2d.)  The full fractal structure
of a strange attractor is only attainable in an unphysical
purely classical limit.  In an actual physical system, the uncertainty
principle provides a lower cutoff to the scale-invariance
of the strange attractor.  It is unphysical to discuss regions of phase
space with areas smaller than Planck's constant.

In fact, the actual lower cutoff is even larger than that, due to the
interaction with the environment.  Classically, one can eliminate noise
from the environment by going to the limit of zero temperature.  In
quantum systems, though, there is noise even in this limit -- the
equivalent of zero point noise for the degrees of freedom of the
reservoir.  The existence of this noise is closely related to the fact
that in general, coarse-grainings much coarser than would be na\"\i vely
suggested by the uncertainty principle are required for decoherence
\cite{Brun2}; in general, phase space cells large compared to $\hbar$
are required.

\subsection{The linearized theory}

The linearized equation
are expected to remain valid as long as the potential is approximately
locally quadratic, i.e., as $\beta\rightarrow0$.
This approximation takes advantage of QSD's unique localization properties.
We can solve these equations,
suitably modified by the presence of the ansatz in (\ref{hamiltonian}),
for different values of $\beta$, and compare them to results generated
with the full QSD code.  The results are displayed in figures 3a--3d.

We see that at $\beta=1$ the linearized
equations are completely wrong (figure 3a).
As we reduce $\beta$, however, the
two methods come into better and better agreement, until at $\beta=0.01$
the agreement is nearly perfect (figures 3b--3d).

The linearized equations are thus only valid when one goes very far
toward the quasiclassical limit.  In this regime, however, the
computational advantage over the full equations is enormous;
for $\beta = 0.01$ it is $10^4$.  In the
intermediate regime of $\beta \sim 0.1$,
the linearized approximation is no longer
valid, but localization can still provide a tremendous advantage.  Using
the MQSD algorithm, one can represent the system by 10--20
basis states, as opposed to the $\sim 1000$ basis states which would
be needed by QSD without the moving basis algorithm and more than
$10^6$ real numbers which would be required to
solve the master equation.

\section{Conclusions}

Many criteria have been suggested to define chaos in quantum systems.
These differ in their emphasis and domain of application.  But nearly
all are based on the premise that, as one approaches the classical realm,
quantum systems should behave more and more like classical chaotic systems.
This limit is difficult in the closed systems which are usually studied.
By contrast, for open systems the classical limit arises naturally.

Quantum state diffusion provides a vivid picture of how this crossover from
quantum to classical occurs.  As a quantum system becomes macroscopic, it
increasingly resembles a localized wave packet, following an approximately
classical path in phase space.  At short length scales, its evolution is
dominated by quantum mechanical uncertainties, due to quantum spreading
and the random influence of the environment.  As one approaches the classical
limit, these stochastic influences become less important, being replaced
by the uncertainties of classical chaos.

This is clearly illustrated by the numerical results of QSD simulations.
These demonstrate the value of
QSD as a practical algorithm for solving the master equation.  By making
use of QSD's unique localization properties, the equations can be solved
numerically with far greater efficiency than would be possible in solving
the master equation directly \cite{Schack1}.

Going to the classical limit this advantage is even more pronounced;
one may replace the full QSD equations by a linearized approximation,
requiring the solution of only 5 coupled ordinary differential equations
\cite{HallZoup}.
This approximation provides not only a powerful numerical tool, but also
demonstrates how quasiclassical equations of motion rise from the
underlying quantum theory.

QSD also provides a practical method of calculating results in the decoherent
histories formalism of quantum mechanics.  While decoherent histories
provides a powerful interpretive tool, it does not necessarily lend
itself to practical calculations; it is often difficult to solve
for the probabilities of individual histories.
For open systems, QSD generates such
histories automatically, with their correct probabilities.
The arguments of Gell-Mann and Hartle, that
probabilities will peak about quasiclassical histories in the classical
limit, are supported by the QSD analysis of this limit, and by the
numerical results presented in this paper.

\section*{Acknowledgments}

We thank the EPSRC in the UK for
financial support, and gratefully acknowledge M Gell-Mann, N Gisin,
J Halliwell, J Hartle, S Lloyd, J Paz, J Ralph, M Rigo, T Spiller, W Strunz,
and W Zurek for many valuable communications.  TAB performed some of this
work while visiting the University of Geneva.

\vfil\eject

\vfil\eject

\vfil

Figure 1.  The constant phase surface of section for the classical forced,
damped Duffing oscillator in the chaotic regime, $\Gamma = 0.125$, $g = 0.3$.

\vfil

Figure 2.  The constant phase surface of section for a single QSD
trajectory of the quantum forced,
damped Duffing oscillator in the chaotic regime, $\Gamma = 0.125$, $g = 0.3$,
for four scalings:  a) $\beta = 1.0$, b) $\beta = 0.25$, c) $\beta = 0.1$,
d) $\beta = 0.01$.

\vfil

Figure 3.  The constant phase surface of section for a single
linearized QSD trajectory of the quantum forced,
damped Duffing oscillator in the chaotic regime, $\Gamma = 0.125$, $g = 0.3$,
for four scalings:  a) $\beta = 1.0$, b) $\beta = 0.25$, c) $\beta = 0.1$,
d) $\beta = 0.01$.

\vfil\end{document}